\documentclass[showpacs,twocolumn,preprintnumbers,amsmath,amssymb]{revtex4} 
\usepackage{graphicx}  
\usepackage{dcolumn}   
\usepackage{bm}        
\usepackage{amssymb}   
\usepackage{hyperref}
\usepackage{xspace}
\usepackage{amsmath}
\usepackage{lineno}


\begin{document}

\widetext
\preprint{CERN-PH-EP-2011-228} 
\preprint{Submitted to Physical Review Letters}
\vspace*{0.1cm}
\title{Search for a light Higgs boson decaying to long-lived weakly-interacting particles in proton-proton collisions at $\sqrt{s}$ = 7 TeV with the ATLAS detector}
\author{The ATLAS Collaboration}

\begin{abstract}
 A search for the decay of a light Higgs (120~$-$~140~GeV) to a pair of weakly-interacting, long-lived particles in 1.94~fb$^{-1}$ of proton-proton collisions at $\sqrt{s}$~=7~TeV recorded in 2011 by the ATLAS detector is presented.  The search strategy requires that both long-lived particles decay inside the muon spectrometer.  No excess of events is observed above the expected background and limits on the Higgs boson production times branching ratio to weakly-interacting, long-lived particles are derived as a function of the particle proper decay length.
\end{abstract}

\maketitle

A Higgs boson~\cite{HIGGS1,HIGGS2,HIGGS3} below 140~GeV is particularly sensitive to new physics.  Many extensions of the Standard Model (SM) include neutral, weakly-coupled particles that can be long-lived~\cite{HV2,CDGW} and to which the Higgs may decay. These long-lived particles occur in many models, including gauge-mediated extensions of the Minimal Supersymmetric Standard Model (MSSM)~\cite{GMSB}, MSSM with R-parity violation~\cite{RPARITY}, inelastic dark matter~\cite{IDM} and the Hidden Valley (HV) scenario~\cite{HV1}.

  This letter presents the first ATLAS search for the Higgs decay, \mbox{$h^0\rightarrow\pi_v\pi_v$}, to two identical neutral particles ($\pi_v$) that have a displaced decay to fermion$-$anti-fermion pairs. As a benchmark we take an HV model~\cite{HV1} in which the SM is weakly coupled, by a heavy communicator particle, to a hidden sector that includes a pseudoscalar, the $\pi_v$.  Due to the helicity suppression of pseudoscalar decays to low-mass $f\overline{f}$ pairs, the $\pi_v$ decays predominantly to heavy fermions, $b\overline{b}$, $c\overline{c}$ and $\tau^{+}\tau^{-}$ in the ratio 85:5:8 $\%$.  The weak coupling between the two sectors leads the $\pi_v$ to have a long lifetime. Other, non-HV, models with the identical signature, where the $\pi_v$ is replaced with another weakly-interacting scalar or pseudoscalar particle, are discussed in Ref.~\cite{HV2,CFW}. Both Tevatron experiments, CDF and D0, performed similar searches for displaced decays in their respective tracking volumes, which limited the proper decay length range they could explore to a few hundred millimeters~\cite{HVCDF,HVD0}. 

  In many of these beyond-the-SM scenarios, the lifetime of the neutral states is not specified and can have a very large range. The current search covers a range of expected proper decay lengths extending to about 20~m by exploiting the size and layout of the ATLAS muon spectrometer. Consequently the experimental challenge is to develop signature-driven triggers to select displaced decays throughout the ATLAS detector volume~\cite{PUBNOTE}. 

  This analysis requires both $\pi_v$ decays to occur near the outer radius of the hadronic calorimeter (\mbox{$r$~$\sim$ 4~m}) or in the muon spectrometer (MS). Such decays give a \mbox($\eta,\phi$) cluster of charged and neutral hadrons in the MS. Requiring both $\pi_v$'s to have this decay topology improves background rejection. The analysis uses specialized tracking and vertex reconstruction algorithms, described below, to reconstruct vertices in the MS. The analysis strategy takes advantage of the kinematics of the gluon fusion production mechanism and subsequent two-body decay, \mbox{$h^0\rightarrow\pi_v\pi_v$}, which results in events with back-to-back $\pi_v$'s, by requiring two well separated vertices (\mbox{$\Delta R~\equiv~\sqrt{(\Delta\eta)^{2}+(\Delta\phi)^{2}}~>~2$)~\cite{COORDINATES}} in the MS.
  
  The data used in this analysis were collected in the first half of 2011 with the LHC operating at 7~TeV. Applying beam, detector and data quality requirements resulted in a total integrated luminosity of 1.94~fb$^{-1}$. The integrated luminosity has a relative uncertainty of 3.7$\%$~\cite{LUMI1,LUMI2}.
  
  Signal Monte Carlo (MC) samples were generated using {\tt PYTHIA}~\cite{PYTHIA,COLOR} to simulate gluon fusion production (\mbox{$gg\to h^{0}$}) and decay of the Higgs (\mbox{$h^0\rightarrow\pi_v\pi_v$}).  Four samples were generated: $m_{h^{0}}$~=~120 and 140~GeV and for each $m_{h^{0}}$ two $\pi_v$ masses of 20 and 40~GeV. The predicted Higgs production cross sections~\cite{HiggsXs} are: \mbox{$\sigma$($m_{h^0}$=120~GeV)~=~16.6$^{+3.3}_{-2.5}$~pb} and \mbox{$\sigma$($m_{h^0}$=140~GeV)~=~12.1$^{+2.3}_{-1.8}$~pb}, and the branching ratio for \mbox{$h^0\rightarrow\pi_v\pi_v$} is assumed to be 100$\%$. The response of the ATLAS detector was modeled with GEANT4~\cite{GEANT,ATLSIM}.  The effect of multiple $pp$ collisions occurring during the same bunch crossing (pileup) was simulated by superimposing several minimum bias events on the signal event.  The MC events were weighted so that the pileup in the simulation agrees with pileup conditions found in data.  

  ATLAS is a multipurpose detector~\cite{ATLASTDR} consisting of an inner tracking detector (ID) surrounded by a superconducting solenoid that provides a 2~T field, electromagnetic and hadronic calorimeters and a MS with a toroidal magnetic field. The ID, consisting of silicon pixel and strip detectors and a straw tube tracker, provides precision tracking of charged particles for \mbox{$\mid\eta\mid$~$\leq$~2.5}. The calorimeter system covers \mbox{$\mid\eta\mid$~$\leq$~4.9} and has 9.7 interaction lengths at \mbox{$\eta$~=~0}. The MS consists of a barrel and two forward spectrometers, each with 16 $\phi$ sectors instrumented with detectors for first level triggering and precision tracking detectors for muon momentum measurement. Each spectrometer has three stations along the muon flight path: inner, middle, and outer. In the barrel, the stations are located at radii of $\sim$4.5~m, 7~m and 10~m, while in the forward MS, they are located at \mbox{$\mid$$z$$\mid$~$\sim$~7.5~m}, 14~m and 20~m. This analysis uses muon tracking for \mbox{$\mid\eta\mid$~$\leq$~2.4}, where each station is instrumented with two multilayers of precision tracking chambers, Monitored Drift Tubes (MDTs). It also utilizes Level 1~\cite{L1TRIG} (L1) muon triggering in the barrel MS (\mbox{$\mid\eta\mid$~$\leq$~1}). The trigger chambers are located in the middle and outer stations. The L1 muon trigger requires hits in the middle station to create a low $p_{\textrm{T}}$ muon Region of Interest (RoI) or hits in both the middle and outer stations for a high $p_{\textrm{T}}$ RoI. The muon RoIs have a spacial extent of \mbox{0.2$\times$0.2} in \mbox{$\Delta\eta\times\Delta\phi$} and are limited to two RoIs per sector.
      
   A dedicated, signature-driven trigger, the muon RoI cluster trigger~\cite{PUBNOTE}, was developed to trigger on events with a $\pi_v$ decaying in the MS. It selects events with a cluster of three or more muon RoIs in a \mbox{$\Delta R$~=~0.4} cone in the MS barrel trigger chambers.  This trigger configuration implies that one $\pi_v$ must decay in the barrel spectrometer, while the second $\pi_v$ may decay either in the barrel or the forward spectrometer. With this trigger, it is possible to trigger on $\pi_v$ decays at the outer radius of the hadronic calorimeter and in the MS with high efficiency. The backgrounds of punch-through jets~\cite{PTJET} and muon bremsstrahlung are suppressed by requiring no calorimeter jets with \mbox{$E_{\textrm{T}}$~$\geq$~30~GeV} in a cone of \mbox{$\Delta R$~=~0.7} and no ID tracks with \mbox{$p_{\textrm{T}}$~$\geq$~5~GeV} within a region of \mbox{$\Delta\eta\times\Delta\phi$~=~0.2$\times$0.2} around the RoI cluster center.  These isolation criteria result in a negligible loss in the simulated signal while significantly reducing the backgrounds.  

\begin{figure}[ht!]
    \centering
    \includegraphics[width=\columnwidth]{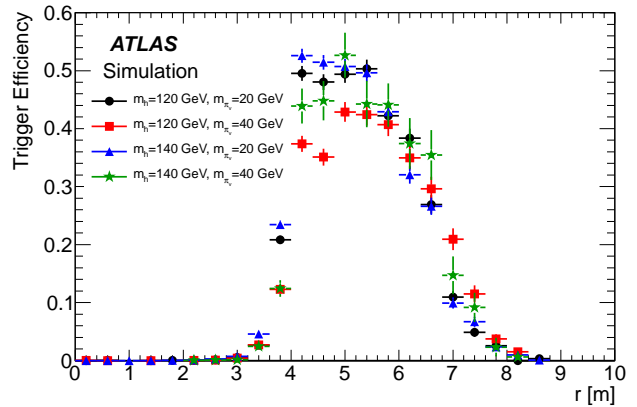}\\
    (a)\\
    \includegraphics[width=\columnwidth]{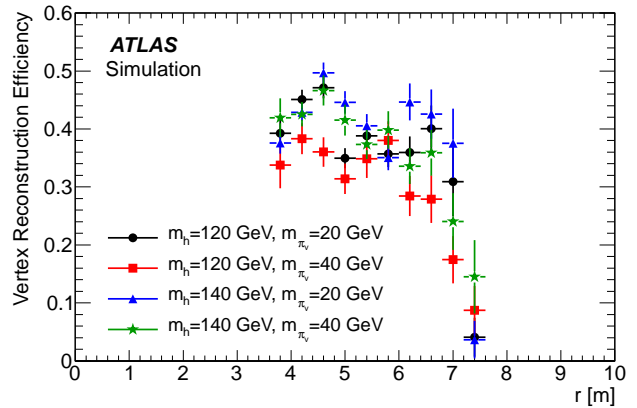}\\
    (b)
    \caption{a: Efficiency of the trigger, as a function of the radial decay position ($r$) of the $\pi_v$.  b: The vertex reconstruction efficiency for $\pi_v$ decays in the barrel for events that pass the muon RoI cluster trigger as a function of the radial decay distance.  The error bars represent the statistical uncertainty on the efficiencies.}
    \label{fig:eff}
  \end{figure}

   As depicted in Fig.~\ref{fig:eff}(a)\cite{efficiencyFLUCT}, MC studies show the RoI cluster trigger is \mbox{$\sim$30~$-$~50$\%$} efficient in the region from 4~m to 7~m.  The $\pi_v$'s that decay beyond a radius of $\sim$7~m do not leave hits in the trigger chambers located at $\sim$7~m, while the $\pi_v$ decays that occur before $r\sim$4~m are located in the calorimeter and do not produce sufficient activity in the MS to pass the muon RoI cluster trigger.  The $m_{h^{0}}$~=~120~GeV and $m_{\pi_{v}}$~=~40~GeV sample has a relatively lower efficiency because the $\pi_v$'s have a lower boost and arrive later at the MS.  As a result the trigger signal may be associated with the incorrect bunch crossing, in which case the event is lost.

\begin{figure}[ht!]
  \centering
  \includegraphics[width=\columnwidth]{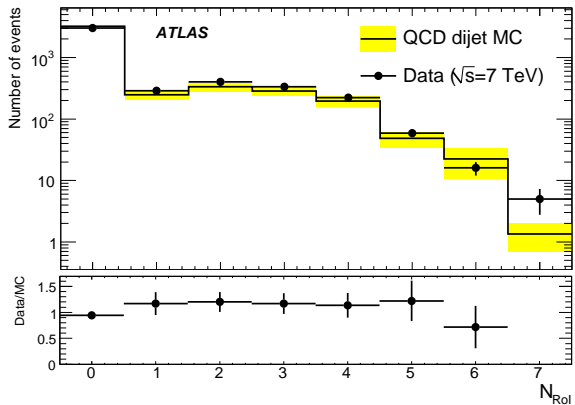}
  \caption{Distribution of number of events vs. number of muon RoIs from punch-through jets contained in the muon RoI cluster for both data and MC events.  The error bands on the QCD dijet MC histogram represent the 1-$\sigma$ statistical uncertainty.}
  \label{fig:ptnroi}
\end{figure}

  The systematic uncertainty of the muon RoI cluster trigger efficiency is evaluated on data using a sample of events containing a punch-through jet. This sample of events is similar to signal events as it contains both low energy photons and charged hadrons in a localized region of the MS.  These punch-through jets are selected to be in the barrel calorimeter (\mbox{$\mid\eta\mid$~$\leq$~1.4}), have \mbox{$E_{\textrm{T}}$~$\geq$~20~GeV}, at least four tracks in the ID, each with \mbox{$p_{\textrm{T}}\geq$~1~GeV} and at least 20~GeV of missing transverse momentum aligned with the jet. To ensure significant activity in the MS, the jet is required to contain at least 300 MDT hits in a cone of $\Delta R$~=~0.6, centered around the jet axis~\cite{singtrkhits}.  The muon RoI cluster trigger algorithm was run in the vicinity of the punch-through jet for both data and MC events. The distribution of RoIs contained in the cluster for data and MC events, normalized to the number of data events, is shown in Fig.~\ref{fig:ptnroi}.  The shapes of the distribution match well between data and MC events.  A horizontal line fit to the ratio, as a function of \mbox{N$_{\textrm{RoI}}$~$\geq$~1}, yields \mbox{1.14~$\pm$~0.09}, and 14$\%$ is taken as the systematic uncertainty. The effects of uncertainties in the jet energy scale (JES)~\cite{JES}, in the initial state radiation (ISR) spectrum~\cite{ISR}, and in the amount of pileup were found to be negligible when varying these quantities by their uncertainties.

  A specialized tracking and vertex reconstruction algorithm was developed to identify $\pi_v$'s that decay inside the MS.  The decay of a $\pi_v$ results in a high multiplicity of low $p_{\textrm{T}}$ particles (\mbox{1~$\leq$~$p_{\textrm{T}}$~$\leq$~5~GeV}) containing $\sim$10 charged particles  and $\sim$5 $\pi^{0}$'s clustered in a small $\Delta R$ region of the spectrometer. The $\pi_v$'s that decay before the last sampling layer of the hadronic calorimeter do not produce a significant number of tracks in the MS.  Thus, detectable decay vertices must be located in the region between the outer radius of the hadronic calorimeter and the middle station of the MS.  Over a wide range of acceptance in the barrel MS, the total amount of material traversed is roughly 1.3 radiation lengths~\cite{ATLASTDR}; therefore, as a consequence of the $\sim$5 $\pi^{0}$'s produced in signal events, large electromagnetic (EM) showers accompany the $\sim$10 charged particles from $\pi_v$ decays. The resulting MS environment contains, on average, approximately 800 MDT hits, of which $\sim$75$\%$ are from the EM showers.
  
  The design of the muon chambers~\cite{ATLASTDR} is exploited in order to reconstruct tracks in this busy environment. The separation of the two multilayers inside a single muon chamber provides a powerful tool for track pattern recognition.  This separation provides enough of a lever arm to allow, in the barrel, a momentum measurement with acceptable resolution for tracks up to approximately 10~GeV~\cite{PRES}. In the forward spectrometers, the muon chambers are outside the magnetic field region; therefore it is not possible to measure the track momentum inside of a single chamber.  In both cases, the tracklets used in the vertex reconstruction are formed using hits in single muon chambers.
  
  The MS vertex algorithm begins by grouping the tracklets using a simple cone algorithm with $\Delta R$~=~0.6.  In the barrel the tracklets are extrapolated through the magnetic field, and the vertex position is reconstructed as the point in ($r$,$z$) that uses the largest number of tracklets to reconstruct a vertex with a $\chi^2$ probability greater than 5$\%$. In the forward spectrometer, the reconstructed tracklets do not have a measurement of the momentum; therefore, the vertex is found using a least squares regression, that assumes the tracklets are straight lines.  Vertices are required to be  reconstructed using at least three tracklets, point back to the Interaction Point (IP)~\cite{VXPOINT} and have \mbox{$\mid\eta\mid$~$\leq$~2.2}.  After requiring the MS vertex to be separated from ID tracks with $p_{\textrm{T}}$~$\geq$~5~GeV and jets with $E_{\textrm{T}}$~$\geq$~15~GeV by \mbox{$\Delta R$~=~0.4} and \mbox{$\Delta R$~=~0.7}, respectively, the algorithm has an efficiency of $\sim$40$\%$ in signal MC events throughout the barrel region (\mbox{4~$\leq$~$r$~$\leq$~7.5~m}) and a resolution of 20~cm in $z$, 32~cm in $r$ and 50~mrad in $\phi$. In the forward spectrometer, the algorithm is $\sim$40$\%$ efficient in the region \mbox{8~$\leq$~$\mid$$z$$\mid$~$\leq$~14~m}.  Fig.~\ref{fig:eff}(b)\cite{efficiencyFLUCT} shows the vertex reconstruction efficiency for the barrel reconstruction algorithm in MC signal events that passed the muon RoI cluster trigger.
 
  The MC description of hadrons and photons in the MS was validated on the same sample of events containing a punch-through jet used to evaluate the trigger performance.  The fraction of these jets that produce a MS vertex was compared in data and QCD dijet MC. Table~\ref{tab:ptvx} shows the fraction of punch-through jets that produce a vertex in data and MC events as a function of the number of MDT hits in a cone of \mbox{$\Delta R$~=~0.6} around the jet axis.  The data-to-MC ratio is fit to a flat distribution that yields a ratio consistent with unity with a 15$\%$ statistical uncertainty, which is taken to be the systematic uncertainty in the vertex reconstruction efficiency.  The systematic uncertainties arising from the JES, ISR spectrum and the amount of pileup were estimated by varying these quantities by their uncertainties and calculating the change in the vertex reconstruction efficiency. The total systematic uncertainty of 16$\%$ for the efficiency of reconstructing a vertex is the sum in quadrature of the uncertainties in the efficiency of the isolation criteria due to varying the JES, ISR and pileup (3$\%$, 3$\%$ and 2$\%$ respectively) and the uncertainty in the comparison of data and MC (15$\%$). 
\begin{table}[ht!]
  \centering
  \begin{tabular}{ccc}
    \hline\hline
    Number of & QCD dijet & Data \\
    MDT hits & Monte Carlo & \\\hline
    300 $\leq$ N$_{\textrm{MDT}}$ $<$ 400 & 10.1$\pm$2.2 $\%$ & 9.1$\pm$0.5 $\%$ \\
    400 $\leq$ N$_{\textrm{MDT}}$ $<$ 500 & 9.2$\pm$2.8 $\%$ & 10.5$\pm$0.7 $\%$\\
    500 $\leq$ N$_{\textrm{MDT}}$ $<$ 600 & 13.1$\pm$5.4 $\%$ & 13.0$\pm$0.9 $\%$\\
    N$_{\textrm{MDT}}$ $\geq$ 600 & 16.5$\pm$4.5 $\%$ & 16.7$\pm$0.7 $\%$\\\hline\hline
  \end{tabular}
  \caption{Fraction of punch-through jets that have a reconstructed vertex in the muon spectrometer for varying numbers of MDT hits for data and QCD Monte Carlo.}
  \label{tab:ptvx}
\end{table}

  The final event selection requires two good MS vertices separated by $\Delta R$~$>$~2. The background due to events with two jets, both of which punch through the calorimeter, is a negligible contribution to the total background due to the tight isolation criteria applied to each vertex.  The background is calculated using a fully data-driven method by measuring the probability for a random event to contain an MS vertex (P$_{\textrm{vertex}}$) and the probability of reconstructing a vertex given the event passed the RoI cluster trigger (P$_{\textrm{reco}}$).  Because P$_{\textrm{vertex}}$ and P$_{\textrm{reco}}$ are measured in data, they incorporate backgrounds from cosmic showers, beam halo and detector noise. The background is calculated as: 
\begin{equation*}
  \begin{split}
    \textrm{N}_{\textrm{Fake}}\textrm{(2 MS }&\textrm{vertex) = N(MS vertex,1 trig)*P}_{\textrm{vertex}}\\&\textrm{ + N(MS vertex,2 trig)*P}_{\textrm{reco}}  
  \end{split}
\end{equation*}
N(MS vertex,1 trig) is the number of events with a single muon RoI cluster trigger object and an isolated MS vertex.  N(MS vertex, 2 trig) is the number of events with an isolated vertex and a second RoI cluster trigger object.  The first term in the equation is the expected number of background events with one vertex that randomly contain a second vertex. P$_{\textrm{reco}}$ is the probability to reconstruct a vertex given there was an RoI cluster trigger; thus, the second term in the equation is the expected number of events with two RoI clusters that have two vertices in the MS.  P$_{\textrm{vertex}}$ was measured using zero bias data~\cite{ZB} to be \mbox{(9.7$\pm$6.9)$\times$10$^{-7}$}, and P$_{\textrm{reco}}$ was measured using the events that pass the muon RoI cluster trigger to be \mbox{(1.11$\pm$0.01)$\times$10$^{-2}$}.  The expected signal would cause, at most, a relative change in P$_{\textrm{reco}}$ of $\sim$1$\%$. P$_{\textrm{reco}}$ was also measured using a sample of events recorded when there were no collisions. In this sample of non-collision background events, P$_{\textrm{reco}}$ was measured to be \mbox{(7.0$\pm$0.6)$\times$10$^{-3}$}. For calculating the background, the larger value of P$_{\textrm{reco}}$ \mbox{(1.11$\times$10$^{-2}$)} is taken since it gives a conservative estimate of the background. N(MS vertex, 1 trig) and N(MS vertex, 2 trig) are 15543 and 1, respectively.  Therefore, the background is calculated to be \mbox{0.03$\pm$0.02} events.  

\begin{figure}[ht!]
    \centering
    \includegraphics[width=\columnwidth]{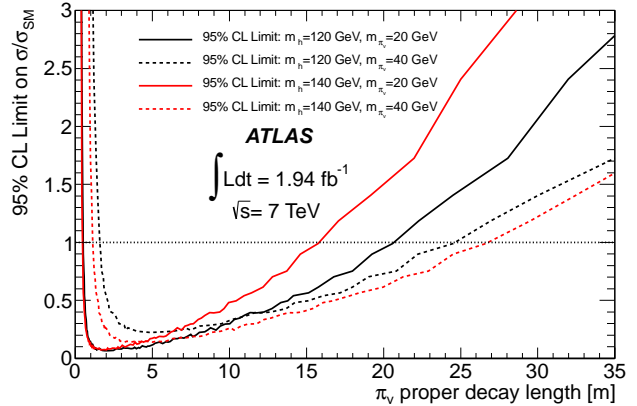}
    \caption{Observed 95$\%$ upper limits on the process \mbox{$h^0\rightarrow\pi_v\pi_v$}, vs. the $\pi_v$ proper decay length, expressed as a multiple of the SM cross section for Higgs production. Exclusion limits assume 100$\%$ branching ratio for the Higgs decaying to $\pi_v$'s.}
    \label{fig:limits}
  \end{figure}

  No events in the data sample pass the selection requiring two isolated, back-to-back vertices in the muon spectrometer. Since no significant excess over the background prediction is found, exclusion limits for \mbox{$\sigma_{h^0}\times{\rm BR}(h^{0}\rightarrow\pi_{v}\pi_{v})$} are set by rejecting the signal hypothesis at the 95$\%$ confidence level using the CLs procedure~\cite{ALREAD}. Figure~\ref{fig:limits} shows the 95$\%$ CL upper limit on \mbox{$\sigma_{h^{0}}\times$BR$(h^0\rightarrow\pi_v\pi_v)$/$\sigma_{\textrm{SM}}$} as a function of the $\pi_v$ proper decay length (c$\tau$) in multiples of the SM Higgs cross section, $\sigma_{\textrm{SM}}$. As expected the Higgs and $\pi_v$ mass combinations with the largest boosts leading to larger $\beta\gamma$c$\tau$ have the smallest exclusion limits.

   In 1.94~fb$^{-1}$ of $pp$ collision data at a center-of-mass energy of 7~TeV there is no evidence of an excess of events containing two isolated, back-to-back vertices in the ATLAS muon spectrometer. Using the model of a light Higgs decaying to weakly-interacting, long-lived pseudoscalars, limits have been placed on the pseudoscalar proper decay length. Table~\ref{tab:limits} shows the broad range of $\pi_v$ proper decay lengths that have been excluded at the 95$\%$ CL, assuming 100$\%$ branching ratio for \mbox{$h^0\rightarrow\pi_v\pi_v$}. These limits also apply to models in which the Higgs decays to a pair of weakly-interacting scalars that in turn decay to heavy quark pairs.   

  \begin{table}[ht!]
    \centering
    \begin{tabular}{ccc}
      \hline\hline
      $m_{h^{0}}$ (GeV) & $m_{\pi_{v}}$ (GeV) & Excluded Region\\\hline
      120 & 20 & 0.50 $<$ c$\tau$ $<$ 20.65 m\\
      120 & 40 & 1.60 $<$ c$\tau$ $<$ 24.65 m\\
      140 & 20 & 0.45 $<$ c$\tau$ $<$ 15.8 m\\
      140 & 40 & 1.10 $<$ c$\tau$ $<$ 26.75 m\\\hline\hline
    \end{tabular}
    \caption{The excluded proper decay lengths (c$\tau$) of the $\pi_v$, at 95$\%$ CL, for each of the signal samples, assuming 100$\%$ branching ratio for the channel \mbox{$h^0\rightarrow\pi_v\pi_v$}.}
    \label{tab:limits}    
  \end{table}

  We thank CERN for the very successful operation of the LHC, as well as the support staff from our institutions without whom ATLAS could not be operated efficiently.  \\
\indent  We acknowledge the support of ANPCyT, Argentina; YerPhI, Armenia; ARC, Australia; BMWF, Austria; ANAS, Azerbaijan; SSTC, Belarus; CNPq and FAPESP, Brazil; NSERC, NRC and CFI, Canada; CERN; CONICYT, Chile; CAS, MOST and NSFC, China; COLCIENCIAS, Colombia; MSMT CR, MPO CR and VSC CR, Czech Republic; DNRF, DNSRC and Lundbeck Foundation, Denmark; ARTEMIS, European Union; IN2P3-CNRS, CEA-DSM/IRFU, France; GNAS, Georgia; BMBF, DFG, HGF, MPG and AvH Foundation, Germany; GSRT, Greece; ISF, MINERVA, GIF, DIP and Benoziyo Center, Israel; INFN, Italy; MEXT and JSPS, Japan; CNRST, Morocco; FOM and NWO, Netherlands; RCN, Norway; MNiSW, Poland; GRICES and FCT, Portugal; MERYS (MECTS), Romania; MES of Russia and ROSATOM, Russian Federation; JINR; MSTD, Serbia; MSSR, Slovakia; ARRS and MVZT, Slovenia; DST/NRF, South Africa; MICINN, Spain; SRC and Wallenberg Foundation, Sweden; SER, SNSF and Cantons of Bern and Geneva, Switzerland; NSC, Taiwan; TAEK, Turkey; STFC, the Royal Society and Leverhulme Trust, United Kingdom; DOE and NSF, United States of America.\\
\indent The crucial computing support from all WLCG partners is acknowledged gratefully, in particular from CERN and the ATLAS Tier-1 facilities at TRIUMF (Canada), NDGF (Denmark, Norway, Sweden), CC-IN2P3 (France), KIT/GridKA (Germany), INFN-CNAF (Italy), NL-T1 (Netherlands), PIC (Spain), ASGC (Taiwan), RAL (UK) and BNL (USA) and in the Tier-2 facilities worldwide.

\onecolumngrid
\clearpage
\onecolumngrid

\begin{flushleft}
{\Large The ATLAS Collaboration}

\bigskip

G.~Aad$^{\rm 48}$,
B.~Abbott$^{\rm 110}$,
J.~Abdallah$^{\rm 11}$,
A.A.~Abdelalim$^{\rm 49}$,
A.~Abdesselam$^{\rm 117}$,
O.~Abdinov$^{\rm 10}$,
B.~Abi$^{\rm 111}$,
M.~Abolins$^{\rm 87}$,
O.S.~AbouZeid$^{\rm 157}$,
H.~Abramowicz$^{\rm 152}$
% ...
\bigskip

% ...
$^{10}$ Institute of Physics, Azerbaijan Academy of Sciences, Baku, Azerbaijan\\
$^{11}$ Institut de F\'isica d'Altes Energies and Departament de F\'isica de la Universitat Aut\`onoma  de Barcelona and ICREA, Barcelona, Spain\\
% ...
$^{48}$ Fakult\"{a}t f\"{u}r Mathematik und Physik, Albert-Ludwigs-Universit\"{a}t, Freiburg i.Br., Germany\\
$^{49}$ Section de Physique, Universit\'e de Gen\`eve, Geneva, Switzerland\\
% ...
$^{87}$ Department of Physics and Astronomy, Michigan State University, East Lansing MI, United States of America\\
% ...
$^{110}$ Homer L. Dodge Department of Physics and Astronomy, University of Oklahoma, Norman OK, United States of America\\
$^{111}$ Department of Physics, Oklahoma State University, Stillwater OK, United States of America\\
% ...
$^{117}$ Department of Physics, Oxford University, Oxford, United Kingdom\\
% ...
$^{152}$ Raymond and Beverly Sackler School of Physics and Astronomy, Tel Aviv University, Tel Aviv, Israel\\
% ...
$^{157}$ Department of Physics, University of Toronto, Toronto ON, Canada\\
% ...
\end{flushleft}



\begin{thebibliography}{}
\bibitem{COORDINATES}
{The ATLAS Collaboration uses a right-handed coordinate system with its origin
  at the nominal interaction point (IP) in the center of the detector and the
  $z$-axis coinciding with the beam pipe. The $x$-axis points from the IP to the
  center of the LHC ring, and the $y$-axis points upward. Cylindrical coordinates
  ($r$,$\phi$) are used in the transverse plane, $\phi$ being the azimuthal angle
  around the beam pipe. The pseudorapidity is defined in terms of the polar
  angle $\theta$ as \mbox{$\eta$ = -ln tan($\theta$/2)}.}
\bibitem{COLOR}
  {The simulation was done in {\tt PYTHIA} 6, with color connections on.  This leads to some quarks from the $\pi_{v}$ decays being connected via a gluon string to partons at the IP.  These events with a macroscopic color string were removed from the MC sample and the remaining events reweighted to preserve the proper branching fractions.}
\bibitem{singtrkhits}
  {For comparison a single, minimum-ionizing track in the barrel MS has about 20 to 25 MDT hits.}
\bibitem{PRES}
  {The momentum resolution for tracklets reconstructed using single MDT chambers in the barrel MS is in the range: $\frac{\delta p}{\mid p\mid}\sim$[0.06-0.09]$\times \mid p\mid$/GeV}.
\bibitem{VXPOINT}
  {The sum of $p_{\textrm{z}}$ of all tracklets used in the vertex fit is required to point back toward the IP.}
\end{thebibliography}
\end{document}